\documentclass[a4paper,fleqn,usenatbib,useAMS]{mnras}

\usepackage{graphicx}	
\usepackage{amsmath}	
\usepackage{amssymb}	
\usepackage{multicol}        
\usepackage{bm}		
\usepackage{pdflscape}	
\usepackage[encapsulated]{CJK}
\usepackage{ucs}
\usepackage[utf8x]{inputenc}

\usepackage[T1]{fontenc}
\usepackage{ae,aecompl}

\usepackage{newtxtext,newtxmath}

\title[Hidden IR structures in NGC\,40]{Hidden IR structures in NGC\,40: signpost of an ancient born-again event}
\author[Toal\'{a} et al.]{J.A.\,Toal\'{a}$^{1}$\thanks{E-mail:\,j.toala@irya.unam.mx}, G.\,Ramos-Larios$^{2}$, M.A.\,Guerrero$^{3}$ and H.\,Todt$^{4}$\\
  $^{1}$Instituto de Radioastronom\'{i}a y Astrof\'{i}sica (IRyA), UNAM Campus Morelia, Apartado postal 3-72, 58090 Morelia, Michoacan, Mexico\\
  $^{2}$Instituto de Astronom\'{i}a y Meteorolog\'{i}a, Depto. de F\'{i}sica, CUCEI, Universidad de Guadalajara, Av. Vallarta 2602, Arcos Vallarta, 44130 Guadalajara, Mexico\\
  $^{3}$Instituto de Astrof\'{i}sica de Andaluc\'{i}a (IAA-CSIC), Glorieta de la Astronom\'{i}a S/N, 18008 Granada, Spain\\
  $^{4}$Institute for Physics and Astronomy, Universit\"{a}t Potsdam, Karl-Liebknecht-Str. 24/25, D-14476 Potsdam, Germany
}

\pubyear{2019}

\begin{document}
\label{firstpage}
\pagerange{\pageref{firstpage}--\pageref{lastpage}}
\maketitle

\begin{abstract}

We present the analysis of infrared (IR) observations of the planetary
nebula NGC\,40 together with spectral analysis of its [WC]-type
central star HD\,826. \emph{Spitzer} IRS observations were used to
produce spectral maps centred at polycyclic aromatic hydrocarbons
(PAH) bands and ionic transitions to compare their spatial
distribution.  The ionic lines show a clumpy distribution of material
around the main cavity of NGC\,40, with the emission from [Ar~{\sc
    ii}] being the most extended, whilst the PAHs show a rather smooth
spatial distribution.  Analysis of ratio maps shows the presence of a
toroidal structure mainly seen in PAH emission, but also detected in a
\emph{Herschel} PACS 70 $\mu$m image. We argue that the toroidal
structure absorbs the UV flux from HD\,826, preventing the nebula to
exhibit lines of high-excitation levels as suggested by previous
authors.  We discuss the origin of this structure and the results from
the spectral analysis of HD\,826 under the scenario of a late thermal
pulse.
  
\end{abstract}

\begin{keywords}
stars: evolution --- stars: winds, outflows --- stars: carbon ---
infrared: ISM --- ISM: molecules --- planetary nebulae: individual:
NGC\,40
\end{keywords}




\section{INTRODUCTION}
\label{sec:intro}

Planetary nebulae (PNe) are formed as a result of the last gasps in the life
of low- and intermediate-mass stars (M$_\mathrm{i} \lesssim$8 M$_{\odot}$).
Previously, these stars have evolved to become asymptotic giant
branch stars (AGB) that eject large amounts of mass into their
circumstellar medium through dense and slow winds
to finally reach the post-AGB phase where they develop fast
stellar winds and strong ionising photon flux.
The wind-wind interaction creates an inner cavity while the UV flux
from the central star photoionises the material, creating a PN
\citep[e.g.,][]{Balick1987}.

\begin{figure*}
\begin{center}
  \includegraphics[angle=0,width=\linewidth]{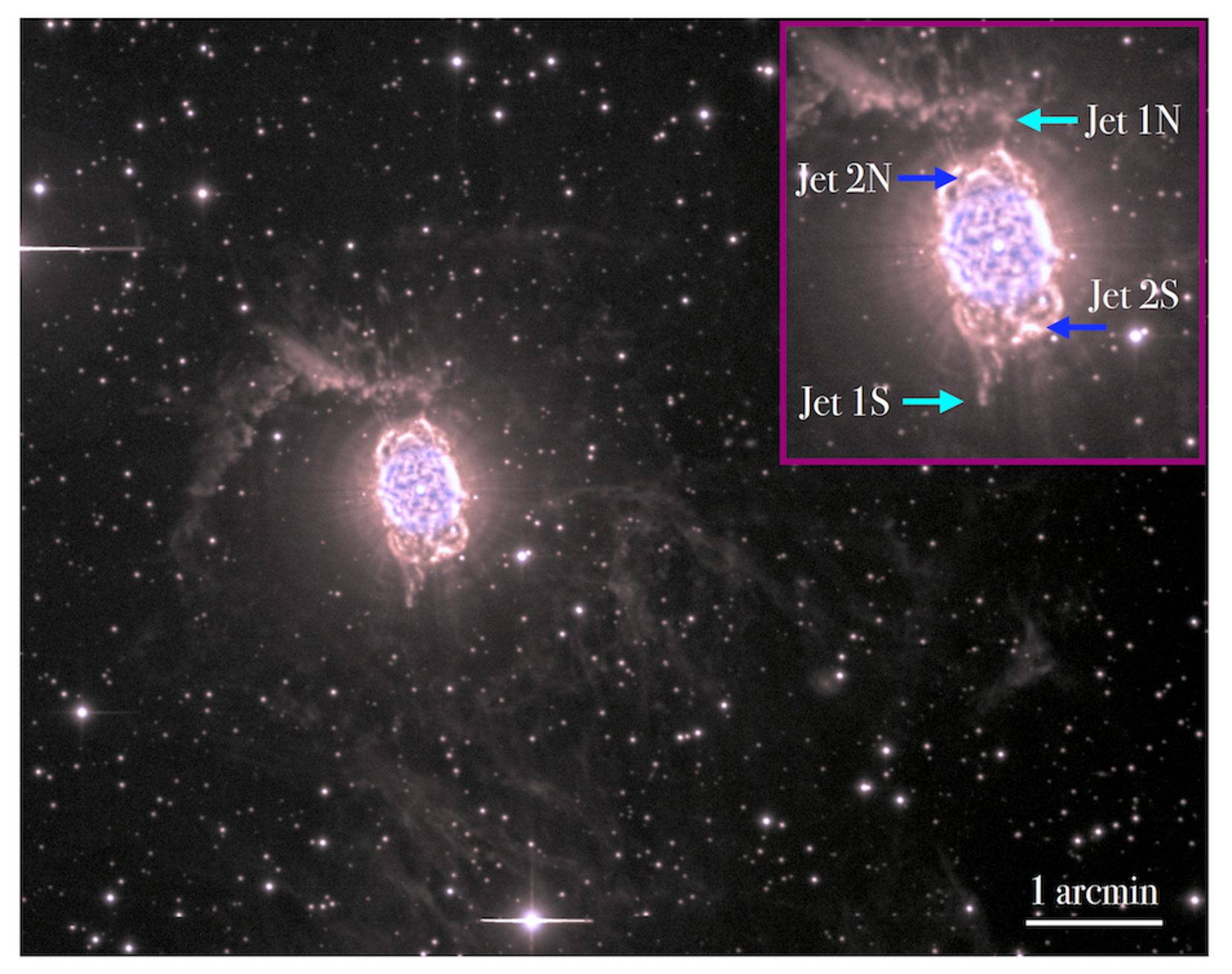}
\label{fig:ngc40_INT}
\caption{Isaac Newton Telescope (INT) Wide-Field Camera (WFC)
  colour-composite picture of NGC\,40.  Red and green correspond to
  the H$\alpha$+[N~{\sc ii}] emission and blue is [O\,{\sc iii}].  The
  inset shows a close-up of the main nebula around HD\,826, the
  central star of NGC\,40, in which we have labelled the main jet-like
  features identified by \citet{Meaburn1996}.  Images courtesy of
  R.L.M.\,Corradi.  North is up, east to the left.  }
\end{center}
\end{figure*}

The PN NGC\,40 around the hydrogen deficient star HD\,826 presents a
variety of morphological features that uncover the different stages of
mass ejection in the formation of PNe.  As shown in the
colour-composite picture presented in Figure~1, the nebula has a
barrel-like main cavity whose northern and southern caps open into a
set of lobes or blisters, sometimes associated with jets, although
there is not kinematical evidence of fast outflows \citep[see,
  e.g.,][]{Meaburn1996}.  A set of optical and IR concentric rings
surround the main cavity \citep{Corradi2004,RamosLarios2011},
disclosing the last gasps in the AGB of HD\,826, the central star of
NGC\,40.  The main nebula is surrounded by an extended filamentary
structure, probably associated with heavy mass-loss events during the
AGB, whose morphology seems to suggest that the nebula is interacting
with the interstellar medium (ISM) due to its relatively high velocity
\citep{Martin2002}.

The nebula exhibits a large C/O ratio, as derived from abundance
determinations based on collisionally excited lines
\citep[e.g.,][]{Pottasch2003}. This is in line with the carbon-rich
      [WC8]-type of its central star
      \citep{Crowther1998}. Furthermore, the \emph{ISO} IR nebular
      spectrum also shows the $\sim$21 and 30~$\mu$m broad emission
      features detected in many C-rich AGB and PNe \citep[see][and
        references
        therein]{Forrest1981,Sloan2014,Mishra2015,Sloan2017}, which
      can be associated with a carbon-rich chemistry, including
      polycyclic aromatic hydrocarbon (PAH) clusters, hydrogenated
      amorphous carbon grains, hydrogenated fullerens, nanodiamonds
      and unidentified carbonaceous material \citep{RamosLarios2011}.
      Detailed modelling of the PAH species in NGC\,40 has been
      recently presented by \citet{Boersma2018}.

Spectral fits to optical and UV spectroscopic observations of HD\,826
based on NLTE models agree that its effective temperature should be
$T_\mathrm{eff} \approx 90$ kK
\citep{Bianchi1987,Leu1996,Marcolino2007}.  However, the nebula
presents low excitation, with the [O~{\sc ii}] $\lambda$~3727 doublet
being the strongest feature in the optical spectrum
\citep[e.g.,][]{Pottasch2003}.  Rough estimates based on the Zanstra
and Stoy methods or even highly sophisticated estimates using
photo-ionisation models point to an effective temperature
$T_\mathrm{eff} \lesssim 40$ kK
\citep{Koppen1978,Preite1983,Harman1966,Pottasch2003,Monteiro2011}.
In order to explain this discrepancy, \citet{Bianchi1987} proposed the
existence of a carbon-rich \emph{curtain} around HD\,826.  These
authors argue that the absorption features seen in the resonance
transition of the C~{\sc ii}~$\lambda\lambda$1334.5,1335.7 doublet
detected in \emph{IUE} observations imply a dense, carbon-rich
circumstellar envelope around HD\,826 expanding at a velocity lower
than its terminal wind velocity
\citep[$V_{\infty}\approx$1000~km~s$^{-1}$;][]{Guerrero2013}.

In this paper we use \emph{Spitzer} IRS observations obtained in map
mode, {\it Herschel} PACS 70~$\mu$m, and Canada-France-Hawaii
Telescope (CFHT) images to assess the spatial distribution of ionic
species, PAHs, H$_2$, and dust in the central region of NGC\,40.  The
comparison of their distributions indeed unveils the presence of
structures inside the main cavity of NGC\,40 close to the central star
that may partially block its UV flux. The paper is organized as
follows.  We describe the observations used here in Section~2 and
present our main results on the spatial distribution of different
components in Section~3. We discuss these results and present a
possible origin for these structures in Section~4.  Finally, a summary
of our findings and conclusions is given in Section~5.

\section{OBSERVATIONS}

We use here \emph{Spitzer} Space Observatory observations of NGC\,40
(program ID 50834, AORKey 29685248; PI: D.\ Weedman) obtained on 2009
March 8.  These correspond to Infrared Spectrograph (IRS) observations
performed in map mode using the short-low (SL) slits covering the
spectral range from 5.2 to 14.5 $\mu$m. Up to 160 slit positions were
placed on the main nebular cavity of NGC\,40 (see Figure~2), with
another 160 slit positions for background subtraction.  The
combination of the 160 slits onto NGC\,40 cover an angular area of
$\sim 64'' \times 60''$.  The pixel scale of the SL detector is
1\farcs8\footnote{\url{https://irsa.ipac.caltech.edu/data/SPITZER/docs/irs/irsinstrumenthandbook/}.}.

The IRS data were processed using the CUbe Builder for IRS Spectra
Maps \citep[{\sc cubism};][]{Smith2007}. {\sc cubism} was used to
visualize spectral maps and to obtain spectra from different regions
within NGC\,40, as well as an integrated spectrum.  Standard
characterization of noise, background subtraction, and bad bixel
removal procedures were followed.

\begin{figure}
\begin{center}
\includegraphics[angle=0,width=1\linewidth]{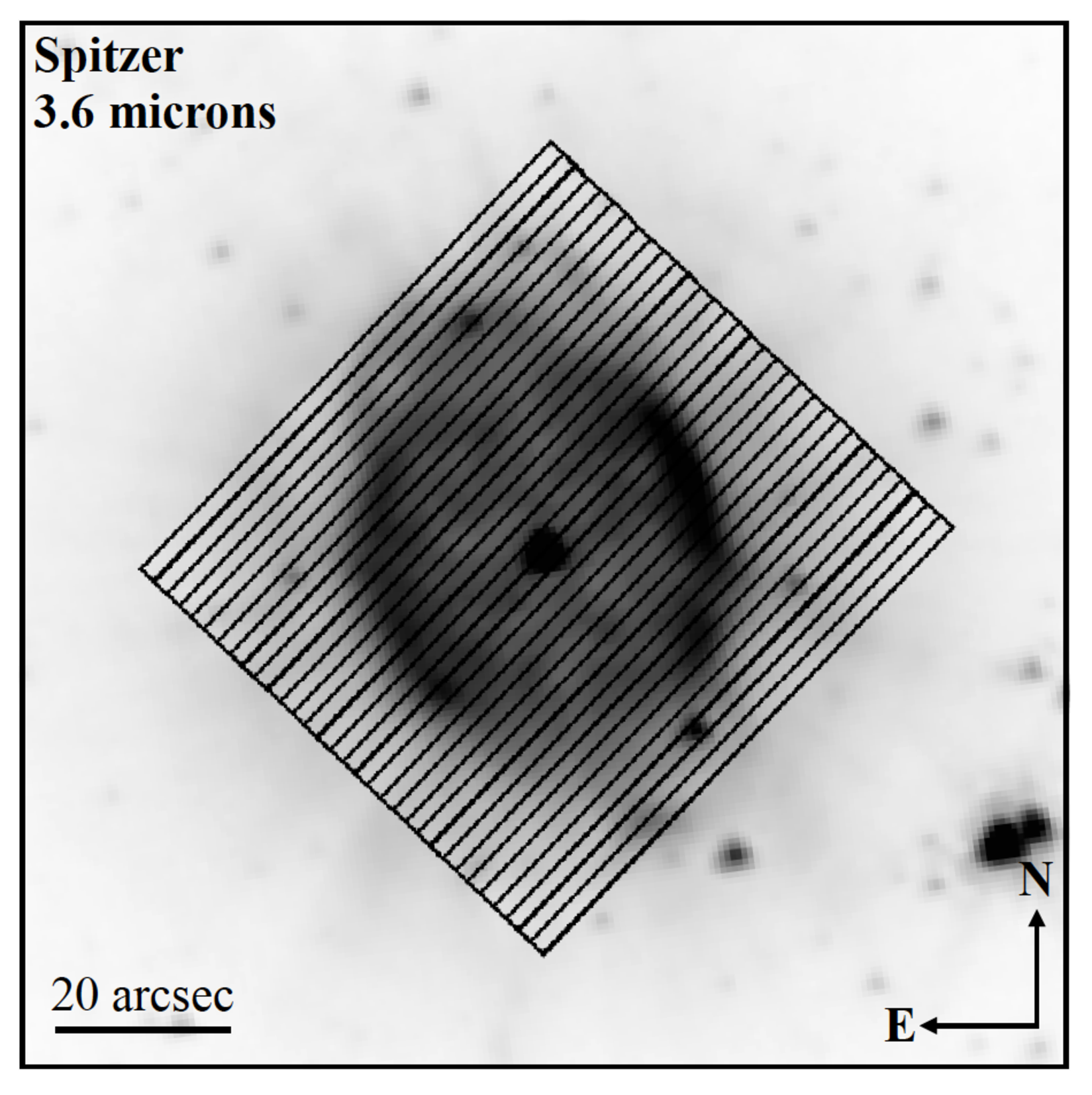}
\label{fig:slits}
\caption{
Slit coverage of the \emph{Spitzer} IRS observations of NGC\,40
overplotted on the \emph{Spitzer} IRAC 3.6 $\mu$m image.
}
\end{center}
\end{figure}

To supplement the \emph{Spitzer} spectral maps of NGC\,40, we also
analysed \emph{Spitzer} IRAC and \emph{Herschel} PACS observations.
The \emph{Spitzer} IRAC observations correspond to the program ID
40115 (AORKey 21976576; PI: G.\ Fazzio) obtained on 2007 October 17,
whereas the \emph{Herschel} PACS data correspond to the observation ID
1342223905 (PI: T.\ Ueta) obtained on 2011 July 10.  All observations
were downloaded from the NASA/IPAC Infrared Science
Archive\footnote{\url{http://irsa.ipac.caltech.edu/Missions/spitzer.html}}.
We have also used near-IR CFHT observations of NGC\,40 obtained with
the Wide-field InfraRed Camera (WIRCam) through the $K$ broadband
($\lambda_\mathrm{c}$=2.146 $\mu$m, $\Delta\lambda$=0.325 $\mu$m) and
H$_2$ 1-0 narrowband ($\lambda_\mathrm{c}=$2.122 $\mu$m,
$\Delta\lambda$=0.032 $\mu$m) filters on 2008 July 18 (PI:
P.\ Martin).

We also obtained UV and optical observations of HD\,826 to model its
spectrum and estimate its stellar properties (see Discussion
section). HD\,826 was observed by the \emph{Far Ultraviolet
  Spectroscopic Explorer (FUSE)} and the \emph{International
  Ultraviolet Explorer (IUE)} satellites. Data from these observations
have been retrieved from MAST, the Multimission Archive at the Space
Telescope Science Institute\footnote{STScI is operated by the
  Association of Universities for Research in Astronomy, Inc., under
  NASA contract NAS5-26555.}. We used the \emph{FUSE} observations
with Obs.\,ID A08501010000 and A08501020000 (PI: H.\,Dinerstein)
obtained on 2000 December 16 in the spectral range 916-1190\,\AA\,
with total exposure times of 15 and 26\,ks.

For the \emph{IUE} spectral range of 1230-1980\,\AA\ and
1900-3220\,\AA\ only low dispersion spectra were availabe. We used the
\emph{IUE} datasets with Obs.\,ID SWP03074 and LWR02656, both taken
with the large aperture with total exposure times of 480\,s for each
pointing. Finally, we obtained high-resolution FIbre-fed Echelle
Spectrograph (FIES) observations at the Nordic Optical Telescope
(NOT). The observations were obtained on 2015 December 11 on service
mode using the high-resolution fibre, which gives a resolution of
$R=67000$ for the 3640--9110~\AA\, wavelength range. The total FIES
exposure time was 900~s.

\section{RESULTS}

Figure~3 presents the integrated background-subtracted {\it Spitzer}
IRS spectrum of NGC\,40.  The spectrum was build adding all the
emission coming from the nebular regions covered by the IRS slits as
shown in Fig.~2.  The spectrum exhibits the clear presence of PAHs at
6.2, 7.7, 8.6 and 11.3 $\mu$m, as well as a small contribution of the
12.0 $\mu$m feature \citep[see also figure~2 in][]{Boersma2018}.  The
spectrum shows notable emission features from low ionisation species
such as [Ar~{\sc ii}] at 6.98 $\mu$m, [Ar~{\sc iii}] at 8.99 $\mu$m,
and [Ne~{\sc ii}] at 12.81 $\mu$m.  A few weak H~{\sc i} emission
lines are also present.  In agreement with \citet{RamosLarios2011}, no
H$_2$ emission lines are detected in the 5--14 $\mu$m spectral range.

\begin{figure}
\begin{center}
  \includegraphics[angle=0,width=\linewidth]{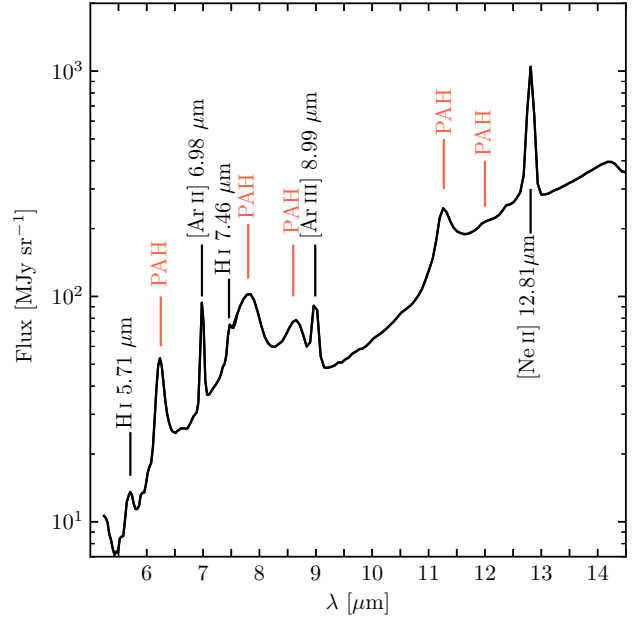}
\label{fig:spec}
\caption{
\emph{Spitzer} IRS low-resolution background-subtracted spectrum of NGC\,40.
The spectrum includes emission from all the SL slits shown in Fig.~2.
Broad line emission attributed to PAHs are marked at 6.3, 7.7, 8.6, 11.3,
and 12.0~$\mu$m.
Other prominent emission lines from ionised species are marked.  
}
\end{center}
\end{figure}

\begin{figure*}
  \includegraphics[angle=0,width=\linewidth]{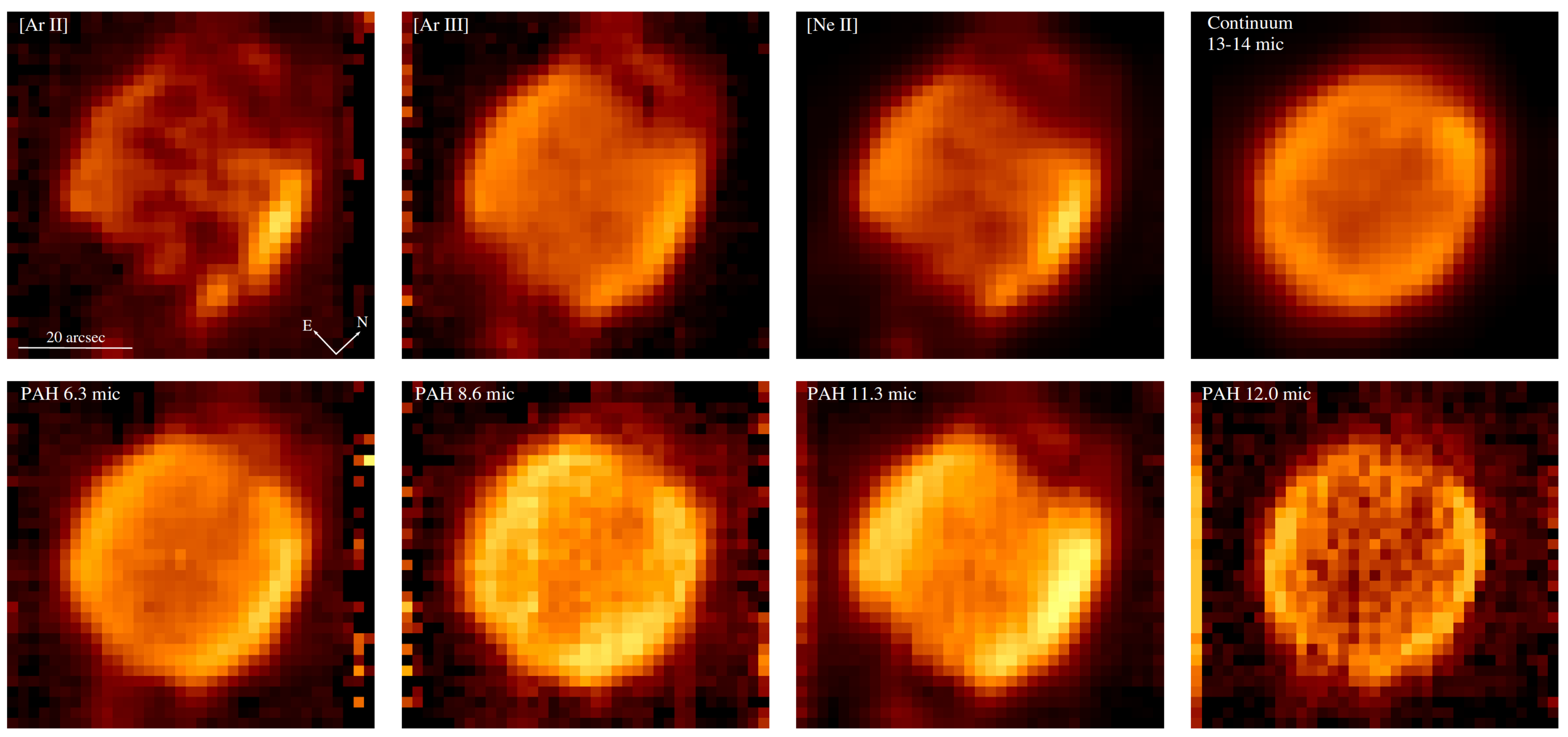}
\label{fig:spec_maps}
\caption{
Spatial maps extracted for different spectral features.
The top-right panel shows a continuum image corresponding to the
emission-free 13--14 $\mu$m spectral range.
All panels have the same field of view and orientation.
}
\end{figure*}

The {\it Spitzer} IRS observations presented here provide us the
opportunity to study the spatial distribution of the emission from
the different PAHs and low-ionisation emission lines detected in
the integrated spectrum of NGC\,40.
To perform this study, we used {\sc cubism} to create spectral maps centred
at the lines of the low-ionisation species and broad emission features of PAHs
(see Figure~3).
For all maps, we subtracted the local continuum following the
procedure described in \citet{Smith2007}.
We do not show the spectral map of the 7.7~$\mu$m PAH feature because
it lays at the edge between the two nodes of the SL orders.  
For completeness, we also extracted a continuum image for the wavelength range
between 13--14~$\mu$m, with no contribution from any line or PAH feature.
For simplicity, we labelled this spectral image as ``continuum''.
All spectral maps are shown in Figure~4.

The spectral maps of the ionic species resemble the optical images
presented in Figure~1: a clumpy barrel-like main cavity with two blobs
aligned N-S just outside this cavity, which are spatially coincident
with the optical blobs or blisters.
The [Ar~{\sc ii}], [Ar~{\sc iii}], and [Ne~{\sc ii}] spectral maps
present a limb-brightened morphology along the minor axis of NGC\,40,
with emission peaking towards the west.
In contrast, the continuum image presents an elliptical shape
with a limb-brightened morphology with no apparent contribution
to any other morphological feature.

The spatial distribution of the emission in the PAH spectral maps
shares some of the properties of that of the emission lines from
ionised species, although some differences are noticeable.
The PAH emission also shows a limb-brightned morphology,
with brighter emission as well in the western region for
the 6.3 and 11.3 $\mu$m features, but not for the 8.6 and
12 $\mu$m ones.  
Note, however, that the emission from the PAHs seems more uniform within
the main cavity than that from the ionised emission lines, particularly
for the PAHs at 6.3, 8.6, and 11.3 $\mu$m (see Fig.~4 bottom panels).  
Otherwise, the northen and southern blobs in the PAH spectral maps are located
closer to the inner cavity than those in the ionic species, but those in the
11.3~$\mu$m PAH feature, which are spatially coincident.

To illustrate the distribution of the ionic species, the continuum and
the PAHs features, we present colour-composite images combining
different spectral maps in the top panels of Fig.~5. The top-left
panel shows the comparison of the three ionic spectral maps.  This
image clearly reveals the clumpy morphology of the emission in the
[Ar~{\sc ii}] and [Ne~{\sc ii}] lines, as well as the extent of the
blowouts north and south of the barrel-like main nebula.  The
top-middle panel compares the distribution of the PAHs.  Although the
emission in the 6.3, 8.6, and 11.3 $\mu$m seems distributed rather
uniformingly inside the main cavity, the northern and southern edges
of the barrel-like structure show noticeable differences. The
top-right panel compares the continuum emission with those of the
[Ar~{\sc ii}] emission line and PAH feature at 11.3 $\mu$m.  The dust
continuum dominates the emission towards the northern and southern
blowouts of the main cavity in NGC\,40.

\begin{figure*}
  \begin{center}
  \includegraphics[angle=0,width=\linewidth]{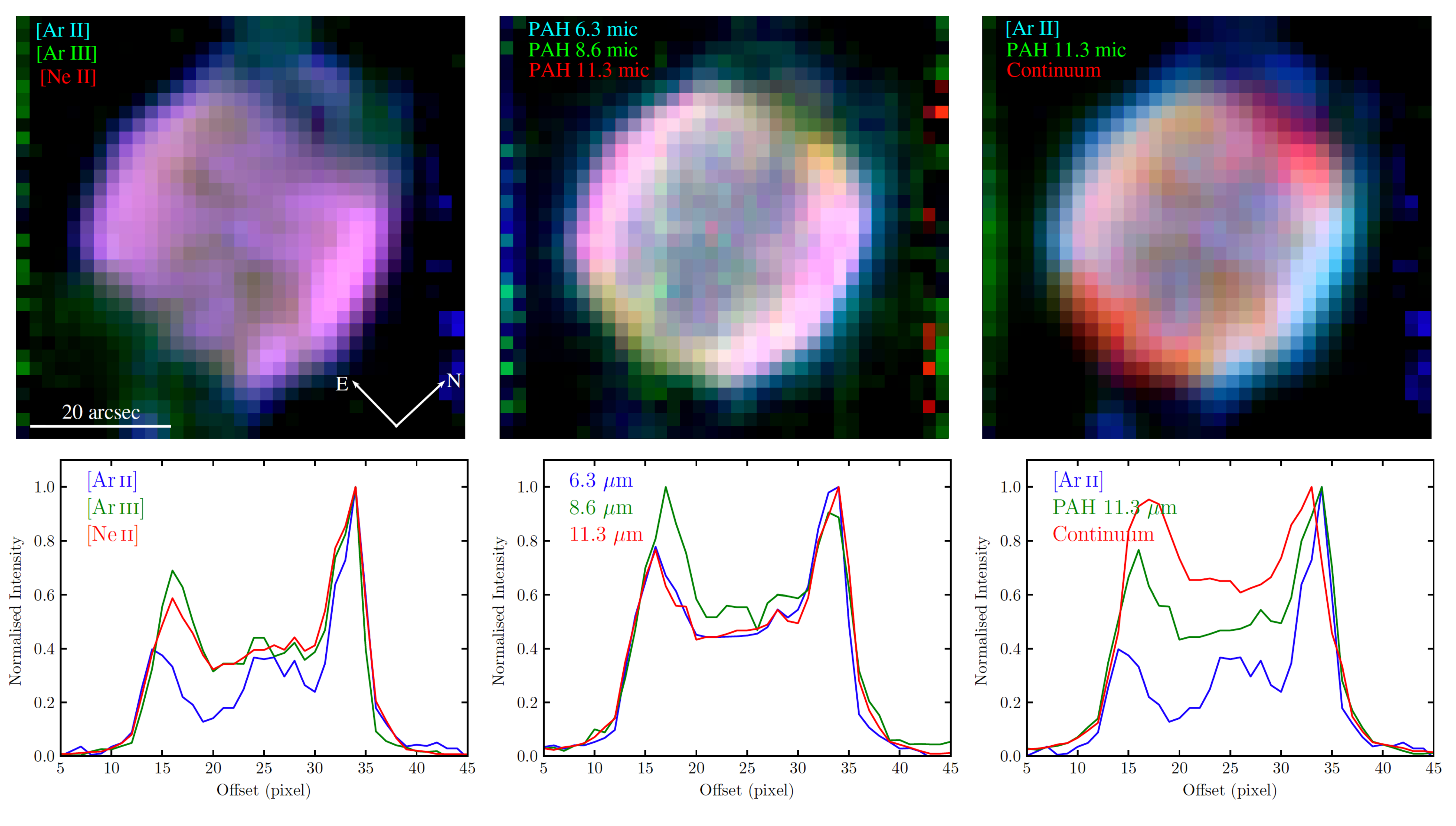}
\label{fig:profiles}
\caption{
The top panels present colour-composite pictures of NGC\,40 combining
different spectral images.
The 
bottom panels show normalised intensity profiles of the different spectral
maps shown in Fig.~4 extracted along the E-W direction (PA=90$^\circ$).
}
\end{center}
\end{figure*}

The bottom row panels in Figure~5 present normalised intensity spatial
profiles extracted from the spectral maps shown in Fig.~4 along the
E-W direction (i.e., PA=90$^\circ$) for those same spectral features
included in the colour-composite pictures of the top row.  The
bottom-left panel confirms that the [Ar~{\sc ii}] emission is more
extended towards the east than that of the [Ar~{\sc iii}] and [Ne~{\sc
    ii}] emission lines. The variations of the [Ar~{\sc ii}] emission
in the inner cavity reflects the clumpy structure unveiled by its
spectral map as shown in Fig.~4. The comparison of the PAH profiles in
the middle-bottom panel show that the PAH at 11.3~$\mu$m is more
extended towards the west and is very similar in shape to that of the
6.3~$\mu$m PAH feature. These profiles generally have a higher valley
to peak ratio than those of the ionised emission lines in the
bottom-left panel. The peaks of the 8.6~$\mu$m PAH lays just inside
the other two profiles. Finally, the bottom-right panel implies the
continuum profile has the highest valley to peak ratio, the peaks are
the broadest, and they peak inside those of the PAH and emission lines
from ionised species. The emission from the 11.3~$\mu$m PAH feature is
very similar in shape to that of the [Ar\,{\sc iii}] emission, which
is likely to be the position of the photodissociated region in
NGC\,40. Moreover, we note that the [Ar\,{\sc ii}] ionic emission
emcompasses that from the PAHs at 11.30~$\mu$m.

\begin{figure*}
\begin{center}
  \includegraphics[angle=0,width=\linewidth]{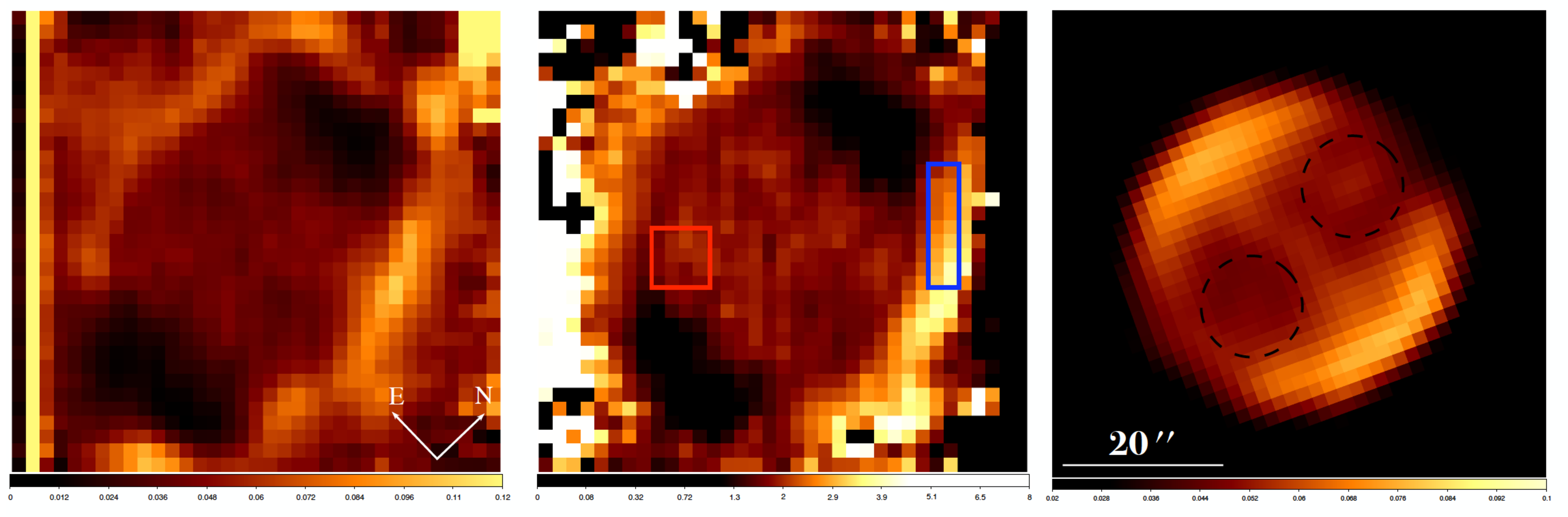}
\label{fig:spec_maps2}
\caption{Spectral map ratios of the PAH 11.3 $\mu$m feature versus the
  continuum (left) and the PAH 6.30 $\mu$m feature (middle). The
  panels have the same field of view and orientation as those in
  Figures~4 and 5. The red and blue boxes overplotted on the middle
  panel indicate the apertures used to extract the torus and rim
  spectra shown in Figure~8. The right panel shows the {\it Herschel}
  PACS 70~$\mu$m image of the same field of view of the other
  panels. The contrast of the {\it Herschel} 70~$\mu$m image has been
  chosen to enhance the structures inside the main cavity in
  NGC\,40. The (dashed-line) circles highlight the location of
    the two blobs of enchanced emission.}
\end{center}
\end{figure*}

Figure~5 hints at the presence of structural components inside the
main cavity of NGC\,40.  In order to highlight this spatial feature,
we investigated different ratio maps using the PAHs, continuum and
ionic species maps.  Interestingly, the PAH ratios unveil a wide
structure inside the main cavity in NGC\,40 extending from east to
west with a width $\sim$30$^{\prime\prime}$. A similar structure can
also be appreciated when comparing the PAHs with the continuum map.
We show in Figure~6 the PAH 11.3/continuum and PAH 11.3/6.3 ratio
images. A dark lane seems to decrease the ratio maps at the middle
plane from east to west\footnote{Similar ratio maps have been created
  using the continuum-subtracted ionic line ([Ar\,{\sc ii}], [Ar\,{\sc
      iii}] and [Ne\,{\sc ii}]) maps in order to search for
  differences in ionisation structure but these images are dominated
  by clumpy emission and the ration maps are not conclusive.}. This
structure is detected in emission in the {\it Herschel} 70~$\mu$m
image presented in Fig.~6 right panel. We note that this structure is
also easily spotted in the PAH ratio maps presented by
\citet{Boersma2018}. In addition to brighter emission at the
equatorial region, this image unveils the presence of two blobs.

\begin{figure*}
\begin{center}
  \includegraphics[angle=0,width=\linewidth]{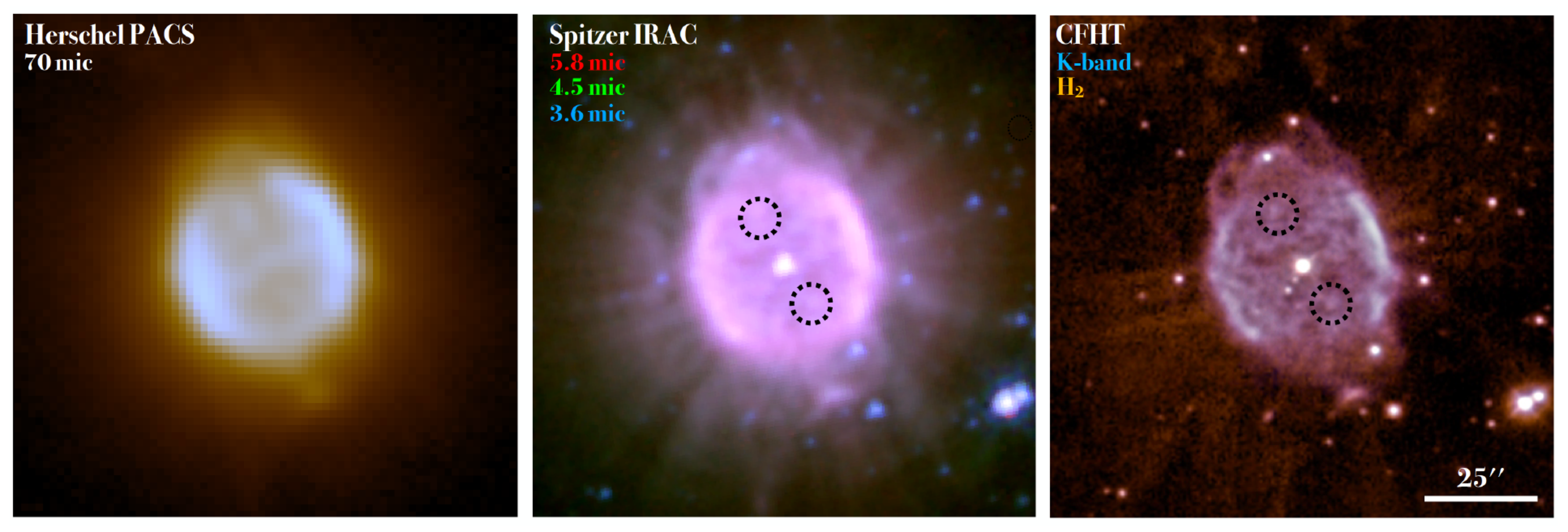}
\label{fig:RGB_images}
\caption{Colour-composite IR images of NGC\,40. Left: {\it Herschel}
  PACS 70~$\mu$m shown at different scales (red - logarithmic, green -
  square root, blue - lineal). Center: {\it Spitzer} IRAC
  image. Right: CFHT WIRCam near-IR image. The circular regions in the
  top-central and -right panels show the position of the two blobs
  unveiled in the {\it Herschel} PACS 70~$\mu$m image (see also Fig.~6
  right panel). The {\it Herschel} and {\it Spitzer} bottom images
  have been processed to enhace the extended emission.}
\end{center}
\end{figure*}

A wide-field image of the {\it Herschel} PACS 70~$\mu$m is presented
in Figure~7 in comparison with other colour-composite IR images of
NGC\,40. It is interesting to note that the CFHT near-IR images
(Fig.~7 right panel) show that H$_2$ is not present in the main
cavity, but its emission can be traced as rays protrunding from it and
along the equatorial (East-West) plane.

A detailed analysis of the IR spectral energy distribution (SED) of
the different morphological features of NGC\,40 would be desirable,
but the available data in the \emph{Herschel} PACS Spectroscopy
Catalogue
\citep{RamosMedina2018}\footnote{\url{https://throes.cab.inta-csic.es/}}
cannot spatially resolve NGC\,40.  The spatial location of continuum
emission and the dominant emission line in the PACS~70~$\mu$m range,
the [O~{\sc i}] 63.2 \AA\, emission line at cannot be determined.
Future SOFIA observations will be used to spatially resolve the
toroidal structure around HD\,826, to construct detailed SED at longer
wavelengths, and to further constrain the properties of these
structures in NGC\,40.

\section{DISCUSSION}

\citet{Boersma2018} reported the clear presence of PAH emission around
HD\,826, the central star of NGC\,40, although they reckon there is no
obvious substructure in their PAH charge maps.  The detailed
comparison of \emph{Spitzer} and \emph{Herschel} mid-IR images and
ratio maps of NGC\,40 shown in Fig.~6 shows it might be interpreted as
a toroidal structure inside its ionised shell.  Whereas its detection
in \emph{Herschel} PACS 70 $\mu$m images is indicative of a dust
component, the \emph{Spitzer} ratio maps reveal an important content
of PAHs, i.e., carbonaceous molecules.  It is worthwhile to emphasize
that the spatial location of PAHs in NGC\,40 does not follow the
typical stratified distribution of material seen in photo-dissociation
regions, where it is associated with dense knots and the presence of
other molecules such as H$_2$ \citep[see the case of NGC\,6720
  presented by][]{Cox2016}.  Neither it compares with the toroidal
structures reported in PNe with dual-dust or mixed chemistry, i.e.,
PNe with carbon- and oxygen-rich dust \citep[see,
  e.g.,][]{GuzmanRamirez2014}.  In those cases, the tori have notable
ionisation gradients, with ionised species in the innermost regions
and PAHs in their outer edges, as the result of the formation of PAHs
by photodissociation of CO \citep{GuzmanRamirez2011}.

There is additional evidence of the presence of large amounts of dust
and carbon-rich material between HD\,826 and the ionised material of
NGC\,40. First, the morphology of this IR feature in NGC\,40 is
similar to the emission map of the C~{\sc iv}
$\lambda\lambda$1548,1551 \AA\ UV emission lines in the image through
the \emph{UVIT} Sapphire filter presented by \citet{Kameswara2018}.
Then, the low ionisation degree of the PAHs in NGC\,40 points to a
high internal attenuation, up to $A_\mathrm{V} \gtrsim 20$ mag
\citep{Boersma2018}.  This is in sharp contrast with the extinction
derived from the Balmer decrement derived from nebular emission lines
or the dip at $\lambda$2200 \AA\ seen in UV spectra of the central
star, which imply $A_\mathrm{V} \approx 1.2$ \citep[see][for a through
  discussion on this issue]{Pottasch2003}.
Finally, we note the different mid-IR spectra of the toroidal
structure and the barrell-shaped nebular rim of NGC\,40
(Figure~8).
The spectrum of the latter flattens at longer wavelengths, whilst that of
the toroidal structure, closer in shape to the total nebular spectrum,
rises up, indicating a larger content of dust and molecules.  This
interpretation confirms the differential spectral results presented by
\citet{Boersma2018}.

\subsection{NLTE analysis of HD\,826}

Such dusty, carbon-rich toroidal structure would absorb the hard UV
flux from the central star of NGC\,40, i.e., it would be the carbon
{\it curtain} proposed by \citet{Bianchi1987} to explain the
discrepancy between the high effective temperature of HD\,826 and the
anomalously low excitation of the nebula around it. 
This discrepancy has been questionned by studies modelling the nebular 
emission, suggesting that the true temperature of HD\,826 is lower, down 
to 38,000 K \citep{Pottasch2003} or 50,000 K \citep{Monteiro2011}. 
To confirm the discrepancy between the effective temperature of HD\,826
and the nebular excitation of NGC\,40, we have analysed public UV
(\emph{FUSE} and \emph{IUE}) and optical FIES NOT spectra of HD\,826
using the updated version of the Potsdam Wolf-Rayet
(PoWR)\footnote{\url{http://www.astro.physik.uni-potsdam.de/PoWR}} 
NLTE code to produce a detailed atmospheric model 
\citep[][]{Grafener2002,Hamann2004}. 
Additional details on the computing methods can be found in 
\citet{Todt_etal2015} and references therein.  
The synthetic spectrum was corrected for interstellar extinction due to dust 
by the reddening law of \cite{seaton1979}, as well as for interstellar line 
absorption for the Lyman series in the UV range.

For the terminal velocity $v_\infty$, we obtained a value of 1000$\pm$100 
km~s$^{-1}$ from the width of the UV P-Cygni line profiles (cf.\ Fig.~9). 
A microturbulence velocity of about 100 km~s$^{-1}$ provides the additional 
line broadening needed to fit the observed emission line profiles. 
The stellar mass was set to a typical value for CSPNe, $M_*$=0.6~M$_\odot$ 
\citep[see e.g.][]{MillerBertolami2007}, though we note that the value of 
$M_*$ has no noticeable influence on the synthetic spectrum.
The density contrast between a wind with clumps and a smooth wind of the same 
mass-loss rate, the so-called micro-clumping parameter $D$, was derived to 
have a value of 10 from the fitting of the extended electron scattering
wings of strong emission lines \citep[e.g.\ ][]{hillier1991,hamann1998}.

It was not possible to fit consistently the \emph{FUSE}, \emph{IUE} and 
optical photometric data.  
While the \emph{IUE} spectra and optical photometry imply a reddening 
of $E_{B-V}=0.41$\,mag, the \emph{FUSE} and \emph{IUE} spectra suggest 
a higher reddening of 0.6~mag and a higher stellar luminosity of 
$\log(L_*/L_\odot)=4.45$ that would overestimate the optical flux.  
The \emph{IUE} observations from different epochs were checked for 
variability, but no significant differences were found for the well 
exposed SWP03075LL (1978-10-20), SWP19081RL (1983-01-25), and SWP53124LL
(1994-12-19) observations.  
Other \emph{IUE} observations seemed to be underexposed or not centred 
on the central star.  
Similar issues may affect some of the \emph{FUSE} observations, thus we rely 
on the well-exposed \emph{IUE} observations and optical photometry to adopt 
an $E_{B-V}$ of 0.41 mag and $\log(L_*/L_\odot)=3.85$.  
This corresponds to $A_\mathrm{V}$=1.27, similar to what was obtained by 
\citet{Pottasch2003}.

The best-fit parameters of our NLTE model of HD\,826 are listed in Table~1. 
The stellar temperature can be constrained by the relative strengths of the 
emission lines of different ions of the same element, e.g., He\,{\sc i} vs.\ 
He\,{\sc ii} or C\,{\sc iii} vs.\ C\,{\sc iv}.  
For a given stellar temperature and chemical composition, the equivalent 
width of any emission line is largely determined by the ratio between the 
volume emission measure of the wind and the area of the stellar surface, 
which can be expressed in terms of the transformed radius $R_\text{t}$ 
\begin{equation}
  R_\text{t} =
  R_\ast\left[\left. \frac{v_\infty}{2500\,\text{km}\,\text{s}^{-1}}
  \right/ \frac{\dot{M}\sqrt{D}}{10^{-4}\,\text{M}_\odot
  \,\text{yr}^{-1}} \right]^{2/3}~~,
  \label{eq:transradius}
\end{equation}
originally introduced by \citet*{schmutz1989}. In this equation $D$ is
the micro-clumping parameter which is defined as the density contrast
between wind clumps and a smooth wind of the same mass-loss rate. 

Our best-fit model is compared to the optical and UV spectra in
Fig.~9, while the SED of HD\,826 is shown in Fig.~10.  The resultant
effective temperature of our best-fit, $T_\mathrm{eff}$=70.8~kK
\cite[as also derived by][]{Marcolino2007}, confirms the hot status of
HD\,826 and our estimate for the Zanstra temperature is 45~kK.
Although \cite{Crowther2003} determined a higher temperature of
$90\,$kK, we find that such temperature results in C{\,\sc iii}
emission lines much weak than those observed.  Moreover, the
\cite{Leu1996} models result in a too strong iron-forest between
$1600-2000\,$\AA\ and too strong C{\,\sc iii} emission lines.

\cite{Leu1996} and \cite{Marcolino2007} reported a carbon mass-fraction of 
50\% in the atmosphere of HD\,826, whereas \cite{Crowther2003} found it to 
be lower, $\simeq$40\%.  
We tested different carbon mass-fractions and found them to range between 
30\% and 50\% are consistent with the observations, as most lines are not 
very sensitive to the carbon mass-fraction relative to the overall quality 
of the fit.
The diagnostic line-pair He{\,\sc ii}\,5411\,/\,C{\,\sc iv}\,5471 implies a 
C:He abundance ratio of 30:67, but the strength of the C{\,\sc iii}\,5696 
emission line suggests a higher carbon abundance of 50\%.  
We conclude that a C:He abundance ratio of 40:57 gives a good compromise
fit. 
A solar value for $X_\text{N}$ gives a good fit to the nitrogen lines in the 
high resolution FIES optical spectrum, while a nitrogen abundance of
$1.4\times10^{-3}$ by mass results in too strong emission lines. 
These results confirm \cite{Marcolino2007} estimate for the nitrogen abundance 
of $<0.1\%$.  
The Balmer lines of hydrogen are blended with He\,{\sc ii} lines, thus the 
upper limit of $2\%$ of hydrogen by mass estimated by \cite{Leu1996} cannot 
be confirmed.
As the spectral lines of neon are all blended with other lines, only an upper 
limit for the neon abundance can be provided, which is also not very strict, 
given the poor quality of the \emph{IUE} observations.  
We notice the absence of a strong F{\,\sc vi}\,1140 line, which is compatible 
with a solar fluorine abundance. 
Similarly, phosphorus, sulfur, and silicon solar abundances provide a good 
fit to the P, S, and Si UV and Si optical spectral lines.

\begin{figure}
\begin{center}
  \includegraphics[angle=0,width=\linewidth]{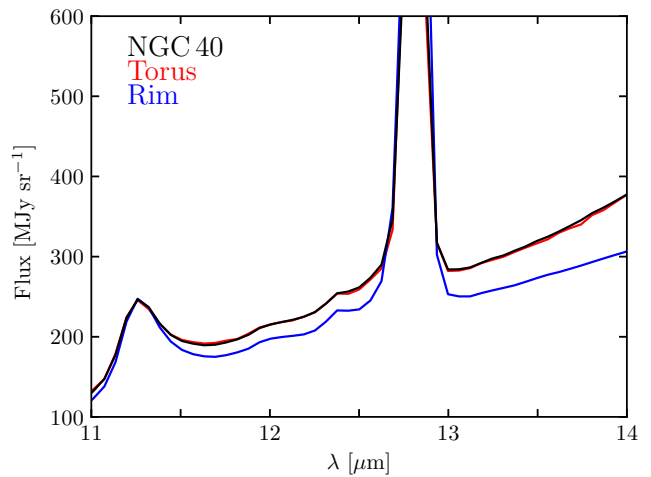}
\label{fig:spec2}
\caption{Comparison of the spectrum extracted from NGC\,40 (see
  Fig.~3) with spectra extracted from the bright rim of NGC\,40 and a
  region spatially coincident with the torus structure shown in Fig.~6
  right panel. The spectra are normalised to the 11.3~$\mu$m feature
  of the NGC\,40 spectrum.}
\end{center}
\end{figure}

\subsection{On the origin of the carbon-rich curtain}

The origin of the inner PAH toroidal structure in NGC\,40 is intriguing.  
Whereas the large nebular C/O ratio \citep{Pottasch2003} suggests that 
the central star underwent the third dredge-up, bringing the resultant 
$^{12}$C to the stellar surface and turning it into a carbon-rich [WC] 
star \citep[][]{Herwig2005}, an eruptive process ejecting carbon-rich 
material inside the ionised nebular shell seems more likely.  
\citet{Herwig2001} discussed several scenarios for the formation of 
[WC]-type central stars of PNe through the {\it late thermal pulse} 
(LTP) event. 
During this event, hydrogen-deficient, carbon-rich material is 
ejected from the central star inside the old hydrogen-rich PN.  
This material coagulates rapidly into carbon-rich dust due to 
the reduction of the effective temperature of the central star. 
Interestingly, the spatial distribution of the hydrogen-poor ejecta detected 
in born-again PNe is generally described as disrupted disks and bipolar
ejections
\citep[e.g.,][]{Borkowki1994,Evans2006}.  

It is thus possible that the central star of NGC\,40 experienced a
LTP, which ejected carbon-rich material preferentially along an
equatorial plane.  As HD\,826 came back to the post-AGB track and
developed a new fast stellar wind, this material formed dust grains
which are now just pushed away into a toroidal structure reaching the
edge of the barrel-like main cavity. Thus, the X-ray emission from
NGC\,40 detected by {\it Chandra} \citep[see][]{Montez2005} might have
an identical origin as that of the born-again PNe A\,30 and A\,78. In
such cases, the carbon-rich material interacts with the
adiabatically-shocked bubble created by the current fast wind,
lowering its temperature to produce diffuse soft X-ray emission
\citep{Guerrero2012,Fang2014,Toala2015}.  Furthermore, the carbon
\emph{curtain} blocks a significant fraction of the UV flux from
HD\,826, resulting in the anomalously low ionisation degree of the
nebular shell. Similar situation has been recently reported in
HuBi\,1, a PN whose central star has experienced a very late thermal
pulse (VLTP) and the carbon-rich dusty ejecta has obscured the star to
a degree such that the outer hydrogen-rich nebular shell has cooled
down and started recombining \citep{Guerrero2018}.
The atmosphere of HD\,826 consists mostly of He and C, with solar nitrogen 
and neon abundances and a likely presence of residual hydrogen.  
These chemical abundances, as expected from stellar evolution models 
\citep[e.g.][]{althaus2005}, favor a LTP event rather than a VLTP one 
\citep{Todt2015}.

\begin{figure*}
\begin{center}
  \includegraphics[angle=0,width=1\linewidth]{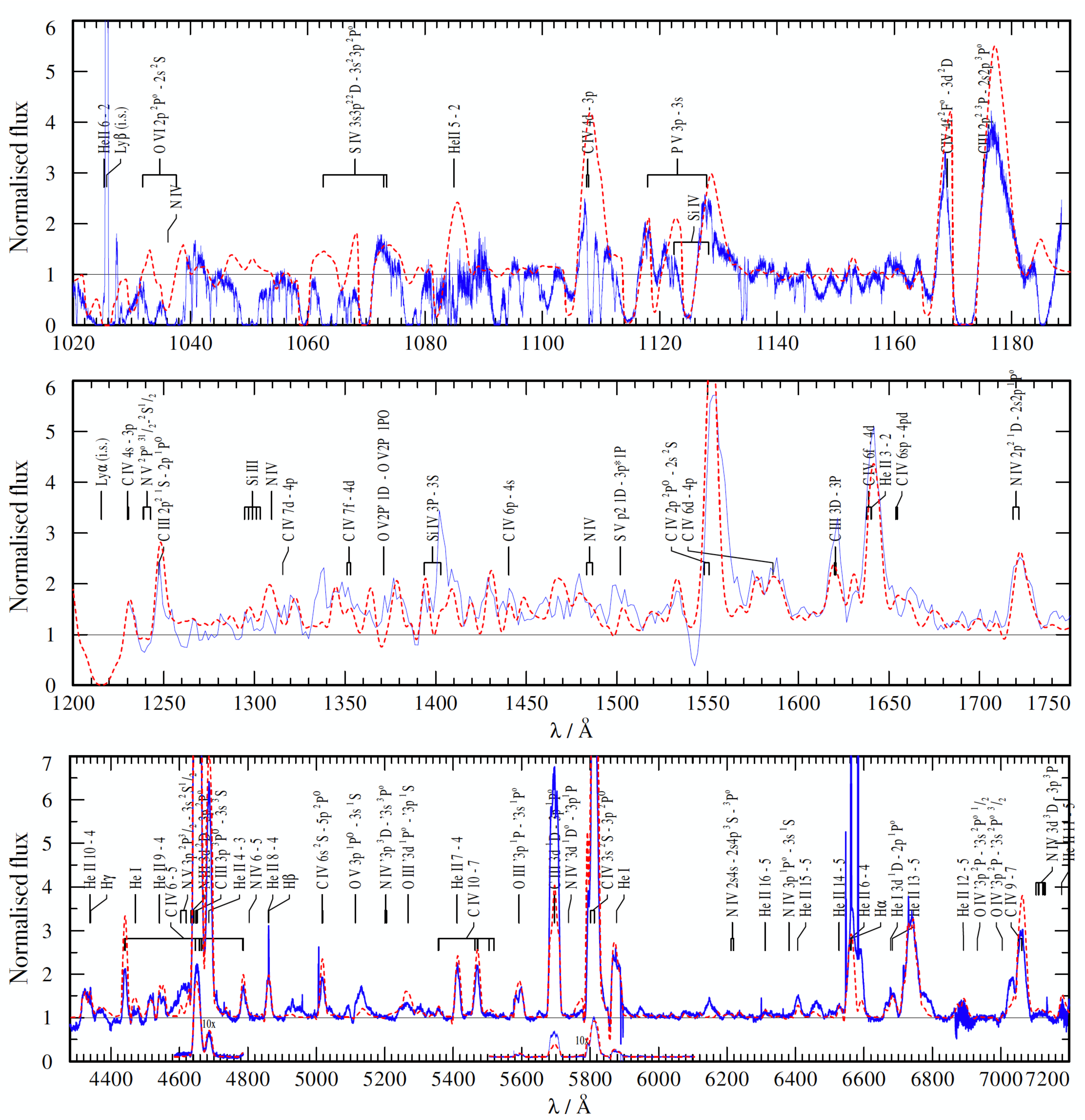}
\label{fig:PoWR_model}
\caption{Details of the best-fit model to HD\,826 obtained with the
  PoWR code. The top panels show a portion of the normalised UV
  spectra ({\it FUSE} and {\it IUE}) whilst the bottom panel shows the
  optical spectrum obtained with FIES at the NOT. The observations are
  shown with blue solid lines and the best-fit model with red dashed
  lines. The {\it FUSE} observation are heavily contaminated by
  interstellar absorption lines, which were not modelled.}
\end{center}
\end{figure*}

\begin{figure*}
\begin{center}
  \includegraphics[angle=0,width=1\linewidth]{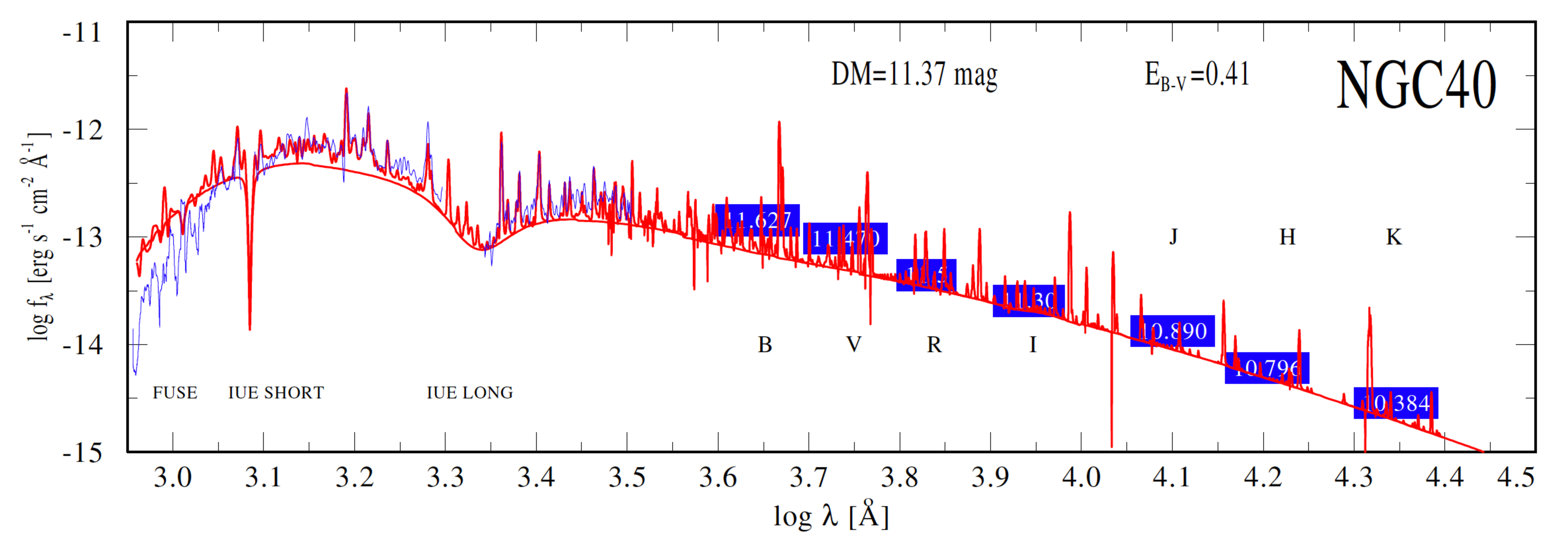}
\label{fig:PoWR_model2}
\caption{Spectral Energy Distribution (SED) of HD\,826 from the UV to
  the IR range. Blue squares are photometric measurements in the
  indicated bands. The SED obtained for our best-fit model is shown in
  red.}
\end{center}
\end{figure*}

\begin{table}
  \caption{Parameters of the central star from analysis with PoWR.}
  \label{tab:parameters}
\begin{tabular}{lcl}
\hline
Parameter & Value & Comment\\
\hline
$T_\text{eff}$ [kK] &  71   &  \\
$d$ [kpc]          &  1.9  &  from \citet{bailerjones2018}\\
$\log(L_*/L_\odot)$  &  3.85 &  \\
$R_*$ [$R_\odot$]  &  0.56 & \\
$R_\text{t}$ [$R_\odot$] & 3.55 \\
$D$                &  10 & density contrast\\
$\log(\dot{M}/\mathrm{M}_\odot\,\text{yr}^{-1})$ & -6.1 & for $D$=10 \\
$v_{\infty}$ [km\,s$^{-1}$] & 1000 & \\
$v_\text{D}$ [km\,s$^{-1}$] & 100 & \\
$M_*$ [M$_\odot$] & 0.6 & adopted \\
$T_\text{Zanstra}$(H\,{\scshape i}) [kK] & 45 \\
$\log Q$(H\,{\scshape i}) & 47.7 & \\
$\log Q$(He\,{\scshape i}) & 46.9 & \\
$\log Q$(He\,{\scshape ii}) & ...$^\dagger$ &  \\
\hline
\multicolumn{3}{c}{Chemical abundances (mass fraction)}\\
\hline
He & $0.57\pm10$ & \\
C  & $0.40\pm10$ & \\
O  & $0.03\pm2$  & \\
N  & $(6.9\pm4) \times 10^{-4}$ & solar \\
Si & $6.7\times 10^{-4}$ & solar\\
P  & $5.8\times 10^{-6}$ & solar \\
S  & $3.1\times 10^{-4}$ & solar \\
F  & $4.6\times 10^{-7}$ & solar \\
Ne & $<0.03$ & \\
H  & $<0.02$ & \\ 
Fe & $1.4\times 10^{-3}$ & iron group elements, solar\\
\hline
\end{tabular}
\\$^\dagger$ no He\,{\sc ii} ionising photon escapes the wind, a black
body model of same temperature and radius would give
log$_{10}Q(\text{He{\,\scshape ii}})=46$. Solar abundances are taken
from \cite{asplund2009}.
\end{table}

Finally, we would like to mention the presence of several pairs of
jet-like ejections in NGC\,40, including those unveiled by the
Herschel PACS 70~$\mu$m image (see Figures 6 - right panel and
Fig.~7). The most extended are those reported by \citet[][labelled as
  Jet\,1 in Fig.~1]{Meaburn1996} and the bipolar structure just
outside the barrel-like main cavity in NGC\,40 (Jet\,2 in
Fig.~1). Fast collimated ejections are a typical signature of
born-again PNe \citep[see, e.g.,][and references
  therein]{Fang2014}. The pair of blobs inside the main cavity of
NGC\,40 (marked in Fig.~7 with dashed-line circular apertures) could
also be interpreted as collimated outflows. Their identification in
optical or near-IR images has been hampered so far due to the dominant
clumpy structure of this PN.

\section{SUMMARY}

We presented an IR study of NGC\,40 around the [WC]-type star HD\,826.
The \emph{Spitzer} IRS low-resolution spectrum of NGC\,40 between 5
and 14 $\mu$m confirms its low level of ionisation, as no emission
lines from high ionisation species are detected in this spectrum. The
main IR spectral features are emission lines from low-ionisation
species, such as [Ar~{\sc ii}], [Ar~{\sc iii}], and [Ne~{\sc ii}], and
PAH clusters at 6.2, 7.7, 8.6, 11.3, and 12.0~$\mu$m.  No H$_2$
emission lines are detected in this spectrum, but the CFHT WIRCam
H$_2$ image unveils the presence of molecular hydrogen just outside
the main cavity.

The analysis of IR observations of NGC\,40 uncovered structures inside
the inner cavity of NGC\,40. The detection of these structures in
\emph{Spitzer} IRS PAH ratio maps and a \emph{Herschel} PACS 70 $\mu$m
image seems to imply a high content of dust and carbon-rich material.
Notably, the structures can be traced in UV C~{\sc iv} images. The
morphology of this emissions suggests that a toroidal structure
surrounds HD\,826.

The current available IR observations suggest that a dusty carbon-rich
toroidal structure is embedded inside the optical nebular shell of
NGC\,40. This toroidal structure around HD\,826 would absorb its UV
flux, as originally suggested by \citet{Bianchi1987}, causing the
nebula to have low excitation. We propose that this structure might
have been originated by a late thermal pulse event experienced by
HD\,826, which also turned it into a [WC]-type star. This would make
NGC\,40 a member of the unpopulated group of born-again PNe.

Future high-resolution mid- and far-IR images and spectra will be
pursuit and will be used to produce a complete modelling of NGC\,40.
This will help characterise these structures and shed light into their
origin.

\section*{Acknowledgements}

The authors are thankful to R.L.M.\,Corradi for providing the INT
nebular images of NGC\,40 and to W.\,Henney for fruitful
discussion. JAT, MAG and HT are funded by UNAM DGAPA PAPIIT project
IA100318. GRL acknowledges support from Fundaci\'on Marcos Moshinsky,
CONACyT and PRODEP (Mexico). MAG acknowledges support of the grant AYA
2014-57280-P, cofunded with FEDER funds. This work has make extensive
use of the NASA's Astrophysics Data System. This work uses public data
from the IR telescopes {\it Spitzer} and {\it Herschel} through the
NASA/IPAC Infrared Science Archive, which is operated by the Jet
Propulsion Laboratory at the California Institute of Technology, under
contract with the National Aeronautics and Space Administration. This
paper also presents data obtained with WIRCam, a joint project of
CFHT, Taiwan, Korea, Canada, France, and the Canada-France-Hawaii
Telescope (CFHT) which is operated by the National Research Council
(NRC) of Canada, the Institut National des Sciences de l'Univers of
the Centre National de la Recherche Scientifique of France, and the
University of Hawaii.





\end{document}